\documentclass{article}
\usepackage[T1]{fontenc}
\usepackage{spconf,amsmath,amssymb,graphicx}
\usepackage[hidelinks]{hyperref}
\usepackage{xcolor}

\title{COMBINING MASKED LANGUAGE MODELING AND CROSS-MODAL CONTRASTIVE LEARNING FOR PROSODY-AWARE TTS}
\name{\begin{tabular}{@{}c@{}}
      Kirill Borodin$^{1}$ \qquad Vasiliy Kudryavtsev$^{1}$ \qquad Maxim Maslov$^{2}$ \\
      Nikita Vasiliev$^{1}$ \qquad Mikhail Gorodnichev$^{3}$ \qquad Grach Mkrtchian$^{2}$
      \end{tabular}}
\address{$^{1}$BitmanagerAI, Dubai, UAE \qquad $^{2}$lab260, Yerevan, Armenia \qquad $^{3}$MTUCI, Moscow, Russia \\
         {\tt kborodin.research@gmail.com}}

\hypersetup{pdftitle={Combining Masked Language Modeling and Cross-Modal Contrastive Learning for Prosody-Aware TTS},
  pdfauthor={Kirill Borodin, Vasiliy Kudryavtsev, Maxim Maslov, Nikita Vasiliev, Mikhail Gorodnichev, Grach Mkrtchian}}
\usepackage{eso-pic}
\AddToShipoutPictureFG*{%
  \put(0,34){\makebox[\paperwidth]{\parbox{6.4in}{\centering\footnotesize
    This work has been submitted to the IEEE ICASSP for possible publication.
    Copyright may be transferred without notice, after which this version
    may no longer be accessible.}}}}

\begin{document}
\ninept
\maketitle
\begin{abstract}
Text encoders in neural text-to-speech (TTS) see only linguistic content, so
prosody must be inferred from paired data and synthesised intonation is often
flat. Contrastive text--audio pretraining (CLAPSpeech) injects prosodic
knowledge by contrasting acoustic realisations of the same phoneme, but it never
trains the encoder to tell different phonemes apart, and its interaction with
masked language modelling (MLM) is unexplored. We study a curriculum for a
speaker-conditioned phoneme-and-subword encoder: MLM, contrastive alignment over
batches of different phonemes, and CLAPSpeech-style same-phoneme refinement. On
Russian conversational speech, with each variant conditioning an identical
Grad-TTS model, MLM followed by mixed-phoneme alignment gives the lowest word
error rate (17.6\% versus 32.8\% for same-phoneme training alone) and the best
quality and intonation ratings. Adding same-phoneme refinement improves
same-phoneme retrieval yet yields the worst synthesis: embedding-space retrieval
does not predict usefulness for generation.
\end{abstract}
\section{Introduction}
\label{sec:intro}
Neural text-to-speech (TTS) systems generate high-fidelity waveforms, yet
listeners still perceive much synthetic speech as flat or inappropriately
intoned. The cause is prosody, i.e.\ the pitch, duration and energy patterns
that are not determined by the phoneme string
alone~\cite{wagner2010prosody,skerryryan2018prosodytransfer}: the same phonemes
are realised differently depending on the surrounding words, the syntax and the
speaker. A TTS text encoder trained only on paired data observes phoneme
identity but has no explicit knowledge of how context shapes prosody, so the
acoustic model regresses towards average
intonation~\cite{huang-etal-2023-prosody,han2025stableformtts}. The problem we
address is therefore: \emph{how should a text encoder be pretrained so that its
per-phoneme embeddings carry context-dependent prosodic information while
remaining a reliable description of what is to be pronounced?}

\textbf{Relation to prior work.} Prosody in TTS has mostly been handled on the
acoustic side: reference encoders and style
tokens~\cite{skerryryan2018prosodytransfer,wang2018styletokens} copy prosody
from an example utterance, and variance predictors~\cite{ren2021fastspeech2}
regress pitch, energy and duration from the text encoding. Both still depend on
how much prosodic information the text encoding itself carries, which is what
text-side pretraining targets. Two pretraining strategies exist.
\emph{Masked reconstruction} learns contextual embeddings by predicting masked
tokens; Prosody-TTS~\cite{huang-etal-2023-prosody} and
DiffProsody~\cite{Oh2024DiffProsody} follow this route but never align text
with audio. \emph{Contrastive alignment}, introduced by
CLAPSpeech~\cite{ye2023clapspeechlearningprosodytext}, trains a text encoder
(phoneme and byte-pair-encoding (BPE) streams) jointly with an audio encoder so
that the text embedding of a token is close to the embedding of the speech
segment that realises it. CLAPSpeech builds every batch from $N$ utterances that
all contain the \emph{same} token (a given phoneme or word) and contrasts that
token across contexts; we call this a \emph{same-token batch}. Because the token
is identical in all $N$ pairs, the only way to match text to audio is through
context-induced prosody. Its authors report lower pitch and duration
prediction errors and higher listener ratings than masked-reconstruction
encoders on three corpora.

Three issues remain open. (i)~In a same-token batch the encoder is never asked
to distinguish one phoneme from another, although a TTS acoustic model relies on
exactly that information; whether a preceding stage with \emph{mixed-phoneme
batches} (each pair targets a different phoneme) is beneficial has not been
tested. (ii)~MLM and contrastive alignment have only been studied separately.
(iii)~CLAPSpeech's text encoder is speaker-agnostic, although prosody is
strongly speaker dependent. Finally, it is unknown whether cheap
embedding-space measures such as text--audio retrieval accuracy can replace
costly TTS training when choosing among pretraining recipes.

We study these issues with a controlled ablation in which every encoder variant
conditions an identical Grad-TTS~\cite{popov2021gradttsdiffusionprobabilisticmodel}
acoustic model trained on Russian conversational speech. We ask:
\textbf{RQ1}~which combination of MLM, mixed-phoneme and same-phoneme
contrastive stages gives the most useful encoder; \textbf{RQ2}~whether
text--audio retrieval accuracy of the encoder predicts synthesis quality; and
\textbf{RQ3}~whether CLAPSpeech-style same-phoneme training helps once the
encoder is already phoneme-discriminative. Our contributions are:
(1)~a speaker-conditioned dual-stream text encoder trained with a pairwise
sigmoid contrastive loss; (2)~evidence that MLM followed by mixed-phoneme
alignment is the best curriculum, reducing word error rate from 32.8\% to
17.6\% and raising intonation ratings from 1.62 to 2.19 relative to same-phoneme
training alone; and (3)~a negative result: appending same-phoneme refinement
improves same-phoneme retrieval but gives the worst synthesis of all variants,
so retrieval scores must not be used as a proxy for generative quality.
\section{Method}
\label{sec:method}

\begin{figure*}[t]
  \centering
  \includegraphics[width=\textwidth]{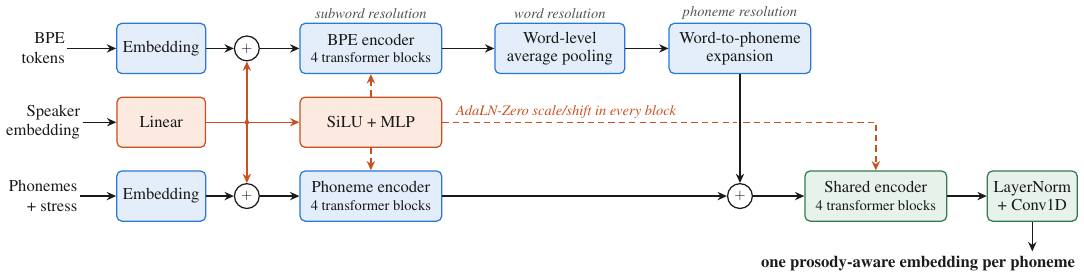}
  \caption{Prosody-aware text encoder. The BPE and phoneme streams are embedded
  and encoded separately; the projected speaker embedding is added to the token
  embeddings and modulates every transformer block through AdaLN-Zero (dashed). BPE states are averaged within each word
  and repeated for every phoneme of that word, summed with the phoneme states,
  and passed through a shared encoder to give one embedding per phoneme.}
  \label{fig:prosody_encoder}
\end{figure*}

\begin{figure}[t]
  \centering
  \includegraphics[width=\columnwidth]{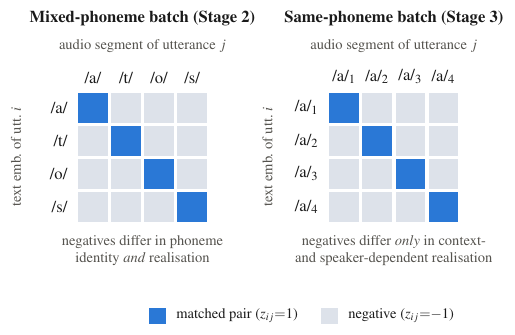}
  \caption{Text--audio similarity matrices of the two batch types (four
  utterances shown). Subscripts denote occurrences of the same phoneme in
  different utterances.}
  \label{fig:batches}
\end{figure}

\subsection{Speaker-conditioned dual-stream encoder}
\label{ssec:arch}
The encoder (Fig.~\ref{fig:prosody_encoder}) maps a sentence to one embedding
per input phoneme symbol. Following
CLAPSpeech~\cite{ye2023clapspeechlearningprosodytext}, it has two input streams:
the phoneme sequence with lexical stress marks, which specifies pronunciation,
and the BPE subword sequence of the same sentence, which carries lexical and
syntactic context. Each stream is processed by four transformer blocks with
relative positional encoding~\cite{shaw_etal_2018_self}, 1-D convolutions in
place of the feed-forward linear layers, and GELU
activations~\cite{hendrycks2017bridging}.

BPE tokens and phonemes segment the sentence differently; the word is the
smallest unit on which the two segmentations agree. We therefore average BPE
states within each word~\cite{ren2022portaspeechportablehighqualitygenerative}
and copy the result to every phoneme of that word. No temporal resolution is
lost, because the phoneme stream is kept at full resolution; the BPE stream
only adds a word-level context vector to it. The summed streams pass through
four further transformer blocks, layer normalisation and a convolutional
projection.

Since prosody is speaker dependent, the encoder is conditioned on a speaker
embedding in two ways: after a linear projection it is added to the token
embeddings of both streams, and it modulates the scale and shift of every
transformer block through AdaLN-Zero~\cite{10377858}. The embedding comes from a
SimAM-ResNet34~\cite{9746294} verification model
(WeSpeaker~\cite{wang2022wespeakerresearchproductionoriented}, trained on
VoxBlink2~\cite{lin24j_interspeech}). At synthesis time it is computed from a
reference utterance of the target speaker other than the one being synthesised.

\subsection{Pretraining stages}
\label{ssec:stages}
The three stages are ordered from general to specific: text-only context
modelling, then grounding of every phoneme in audio, then specialisation on
within-phoneme variation. Fig.~\ref{fig:batches} contrasts the two ways of
building a contrastive batch that distinguish Stages~2 and~3.

\textbf{Stage~1: MLM.} The phoneme and BPE stream encoders are each trained
with BERT-style masked-token
prediction~\cite{devlin2019bertpretrainingdeepbidirectional} (masking
probability 0.15) on text only, giving a contextual initialisation.

\textbf{Stage~2: mixed-phoneme contrastive alignment.} The full encoder is
trained jointly with an ECAPA-TDNN~\cite{Desplanques_2020} audio encoder that
receives the mel-spectrogram. From each of the $N$ utterances in a batch one
phoneme position is drawn at random; let $\mathbf{x}_i$ and $\mathbf{y}_i$ be
the $\ell_2$-normalised text and audio embeddings at that position in utterance
$i$. We minimise the pairwise sigmoid loss of
SigLIP~\cite{zhai2023sigmoidlosslanguageimage},
\begin{equation}
  \mathcal{L} = -\frac{1}{N}\sum_{i=1}^{N}\sum_{j=1}^{N}
  \log\sigma\!\big(z_{ij}\,(t\,\mathbf{x}_i^{\top}\mathbf{y}_j + b)\big),
  \label{eq:siglip}
\end{equation}
where $z_{ij}{=}1$ if $i{=}j$ and $-1$ otherwise, and the scale $t$ and bias $b$
are learned. Unlike the softmax loss of CLAPSpeech, (\ref{eq:siglip}) treats
every pair as an independent binary decision and needs no normalisation over
the batch. Because the drawn positions mostly hold \emph{different} phonemes,
matching text to audio requires both phoneme identity and its acoustic
realisation; we call this a \emph{mixed-phoneme batch}
(Fig.~\ref{fig:batches}, left). Negatives are easy on average, since most
differ in phoneme identity, but roughly one in the size of the phoneme
inventory shares the phoneme and must be told apart by its realisation, so the
stage is not blind to prosody.

\textbf{Stage~3: same-phoneme refinement.} Loss (\ref{eq:siglip}) is kept, but
every batch is a same-token batch in the sense of CLAPSpeech
(Fig.~\ref{fig:batches}, right): all $N$ utterances
contain one chosen phoneme, and $\mathbf{x}_i,\mathbf{y}_i$ are taken at its
position. Phoneme identity no longer separates positives from negatives, so
only context- and speaker-dependent realisation does. Stage~3 used alone is our
re-implementation of the CLAPSpeech phoneme-level training regime within our
architecture and loss; it is the prior-work reference point in all
experiments.
\section{Experimental Setup}
\label{sec:setup}
\textbf{Data.} Contrastive stages and TTS training use the Yandex Podcasts
part of the Balalaika
corpus~\cite{borodin2026balalaika}, a collection of in-the-wild Russian
conversational speech, of which we use about 1{,}000 hours for training. Its annotation pipeline segments recordings with
a semantic voice activity detector, transcribes them with an ensemble of
recognisers combined by ROVER, filters segments by quality and speaker
consistency, and enriches the text with punctuation, normalisation, lexical
stress marks and IPA phonemic transcriptions; phoneme boundaries come from
forced alignment. Conversational podcasts contain far more prosodic variation,
but also more noise and overlap, than read-speech corpora. Audio is
sampled at 22.05\,kHz; following the standard Grad-TTS configuration, the
80-band mel filterbank covers 0--8\,kHz with a hop of 256 samples. Stage~1 uses
the text-only Russian Dialogues corpus~\cite{russian_instructions}.

\textbf{Encoder training.} Stage~1: $7.5{\times}10^4$ steps, batch 512,
learning rate $2.5{\times}10^{-4}$. Stage~2: $2{\times}10^4$ steps, batch 1024,
learning rate $10^{-4}$, $t$ and $b$ initialised to $1/0.07$ and $-10$.
Stage~3: $10^4$ steps, batch 512, learning rate $9{\times}10^{-5}$. A variant
is named by the stages it includes, applied in order (e.g.\ 1+2). Two further
variants of Stage~2 replace (\ref{eq:siglip}) by the symmetric softmax loss of
CLAPSpeech, or remove speaker conditioning.
For the audio side we remove the final pooling and embedding layer of
ECAPA-TDNN, so that it outputs one vector per mel-spectrogram frame;
$\mathbf{y}_i$ is taken from the frames that the forced alignment assigns to
the selected phoneme.

\textbf{TTS.} Each pretrained encoder replaces the text encoder of
Grad-TTS~\cite{popov2021gradttsdiffusionprobabilisticmodel}, which is then
trained for $10^6$ steps (batch 16, learning rate $10^{-4}$) with all other
settings identical. Grad-TTS was chosen because it is cheap enough to train six
times under identical conditions; it is not a state-of-the-art system, and we
interpret its scores only relative to one another (Sec.~\ref{ssec:limits}).

\textbf{Retrieval measures.} For a held-out batch of $N{=}1024$ text--audio
pairs we rank all audio embeddings by cosine similarity to each text embedding;
Recall@$k$ (R@$k$) is the fraction of text embeddings whose own audio segment is
among the top $k$. With a mixed-phoneme batch this gives \emph{R@$k$-diff},
which is mostly solved by phoneme identity and thus measures phoneme
discrimination. With a same-phoneme batch it gives \emph{R@$k$-sim}, which can
only be solved from context-dependent realisation and is our proxy for prosodic
sensitivity. Chance level is $k/N$.

\textbf{Synthesis measures.} We synthesise 200 held-out sentences per system,
the same for all systems. Intelligibility is the word and character error rate
(WER, CER, in \%) of the GigaAM-v2-CTC
recogniser~\cite{kutsakov25_interspeech}; speaker similarity (SIM-o) is the
cosine similarity between SimAM-ResNet100~\cite{9746294} embeddings of the
output and of the reference utterance. As automatic quality predictors we report
UTMOS~\cite{baba2024utmosv2} and the overall NISQA
score~\cite{mittag21_interspeech}.

\textbf{Listening tests.} Since the paper concerns prosody, we ran two separate
tests with native Russian listeners. In the \emph{intonation} test (IntMOS)
listeners judged only how natural the intonation is, on an anchored scale
from 5 (``sounds like a real person, lively'') through 3 (``hard to tell whether
human or synthetic'') to 1 (``clearly robotic'') and 0 (``unintelligible,
intonation cannot be judged''). In the \emph{overall quality} test (MOS) they
rated signal quality on an anchored scale from 5 (``studio quality'') through 3
(``non-professional recording with noticeable noise or distortion'') to 1
(``barely intelligible'') and 0 (``not speech''). These anchored scales are
stricter than the usual unanchored naturalness scale, so absolute values are
not comparable with MOS figures from other papers. Every clip was rated by at
least 7 listeners; following the concerns about listener numbers raised
in~\cite{wester15_interspeech}, we take the per-clip median to limit the
influence of individual raters and report the mean over the 200 clips with a
95\% confidence interval. In total 48 listeners took part; all were unpaid volunteers who gave
informed consent, and ratings were collected with the open-source LabelSpeech
tool~\cite{labelspeech}.
\section{Results}
\label{sec:results}
\begin{table}[t]
  \centering
  \caption{Text--audio retrieval of encoder variants (batch of 1024). -sim:
  same-phoneme batches (prosodic sensitivity); -diff: mixed-phoneme batches
  (phoneme discrimination). $^{\dagger}$CLAPSpeech-style same-phoneme training only. Best in bold, second best underlined.}
  \label{tab:retrieval}
  \setlength{\tabcolsep}{3.5pt}
  \begin{tabular}{l c c c c}
    \hline
    Stages & R@1-sim & R@5-sim & R@1-diff & R@5-diff \\
    \hline
    3$^{\dagger}$ & 0.692 & 0.903 & 0.779 & 0.919 \\
    1+3                  & 0.665 & 0.890 & 0.766 & 0.922 \\
    2                    & 0.690 & 0.911 & \textbf{0.933} & \textbf{0.986} \\
    1+2                  & 0.746 & \underline{0.937} & \underline{0.926} & \underline{0.983} \\
    2+3                  & \textbf{0.782} & \textbf{0.946} & 0.871 & 0.958 \\
    1+2+3                & \underline{0.770} & 0.935 & 0.893 & 0.974 \\
    \hline
    2, softmax loss      & 0.650 & 0.805 & \underline{0.926} & \textbf{0.986} \\
    2, no spk.\ cond.  & 0.529 & 0.825 & 0.871 & 0.979 \\
    \hline
  \end{tabular}
\end{table}
\begin{table*}[t]
  \centering
  \caption{Grad-TTS conditioned on each encoder variant. R@1 values are repeated
  from Table~\ref{tab:retrieval} to show that retrieval does not predict
  synthesis. MOS and IntMOS: mean $\pm$ 95\% confidence interval on the anchored
  0--5 scales of Sec.~\ref{sec:setup}. $^{\dagger}$CLAPSpeech-style same-phoneme training only. Best in bold, second best underlined.}
  \label{tab:tts}
  \setlength{\tabcolsep}{4.5pt}
  \begin{tabular}{l c c c c c c c c c}
    \hline
    & \multicolumn{2}{c}{Encoder retrieval} & \multicolumn{2}{c}{Listening tests}
    & \multicolumn{2}{c}{Predicted quality} & \multicolumn{3}{c}{Intelligibility, speaker} \\
    Stages & R@1-sim & R@1-diff & IntMOS & MOS & UTMOS & NISQA & WER (\%) & CER (\%) & SIM-o \\
    \hline
    3$^{\dagger}$ & 0.692 & 0.779 & $1.615 \pm 0.285$ & $1.605 \pm 0.165$ & 1.43 & 2.28 & 32.8 & 20.8 & 0.798 \\
    1+3   & 0.665 & 0.766 & $1.855 \pm 0.270$ & $1.755 \pm 0.160$ & 1.64 & 2.49 & 31.0 & 15.5 & 0.838 \\
    2     & 0.690 & \textbf{0.933} & $1.890 \pm 0.280$ & $\underline{1.895 \pm 0.190}$ & \underline{1.65} & 2.72 & 29.0 & 14.2 & \underline{0.854} \\
    1+2   & 0.746 & \underline{0.926} & $\mathbf{2.185 \pm 0.255}$ & $\mathbf{1.980 \pm 0.172}$ & 1.62 & \textbf{2.80} & \textbf{17.6} & \textbf{6.7} & \textbf{0.862} \\
    2+3   & \textbf{0.782} & 0.871 & $\underline{2.126 \pm 0.278}$ & $1.840 \pm 0.160$ & \textbf{1.66} & \underline{2.73} & \underline{23.4} & \underline{12.3} & 0.843 \\
    1+2+3 & \underline{0.770} & 0.893 & $1.545 \pm 0.290$ & $1.540 \pm 0.170$ & 1.53 & 2.23 & 42.9 & 30.7 & 0.793 \\
    \hline
  \end{tabular}
\end{table*}

\subsection{RQ1: which curriculum gives the most useful encoder?}
\label{ssec:rq1}
\textbf{Representation level} (Table~\ref{tab:retrieval}). Phoneme
discrimination comes from Stage~2: the variants that include it reach R@1-diff
of 0.87--0.93, whereas same-phoneme training alone (3, 1+3) stays below 0.78,
confirming that same-token batches do not teach the encoder to separate
phonemes. Within Stage~2, the sigmoid loss gives higher prosodic sensitivity
than the softmax loss (R@1-sim 0.690 vs.\ 0.650) at equal discrimination, and
removing speaker conditioning is the most damaging change (R@1-sim 0.529,
R@1-diff 0.871): without the speaker embedding the text side cannot account for
speaker-dependent realisation. MLM initialisation raises R@1-sim of Stage~2
from 0.690 to 0.746.

\textbf{Synthesis level} (Table~\ref{tab:tts}). The 1+2 encoder is best on
every measure except UTMOS, on which the four best variants lie within 0.04.
Relative to the CLAPSpeech-style reference (Stage~3 alone), WER falls from
32.8\% to 17.6\%, SIM-o rises from 0.798 to 0.862, and both listening tests
improve with non-overlapping confidence intervals (IntMOS 1.62 to 2.19; MOS
1.61 to 1.98). The gain in the intonation test is the most direct evidence that
the curriculum improves prosody rather than signal quality only. MLM matters
for synthesis more than Table~\ref{tab:retrieval} suggests: 1+2 and 2 have
almost equal R@1-diff, yet WER drops from 29.0\% to 17.6\%. The effect of MLM is
not uniform, however: it helps before Stage~2 and, slightly, before Stage~3
alone (WER 32.8\% to 31.0\%, IntMOS 1.62 to 1.86), but 1+2+3 is worse than 2+3,
so the benefit of a stage depends on what follows it and stages cannot be
assessed in isolation. Differences between
1+2 and the next-best variants (2+3 on IntMOS, 2 on MOS) are within the
confidence intervals, so we claim superiority of 1+2 over the same-phoneme
reference and over 1+2+3, not over every variant on every measure.

\subsection{RQ2: does retrieval predict synthesis quality?}
\label{ssec:rq2}
\begin{figure}[t]
  \centering
  \includegraphics[width=\columnwidth]{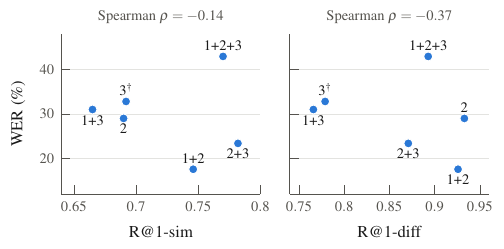}
  \caption{WER of Grad-TTS against the retrieval accuracy of the encoder that
  conditions it; labels give the stages used ($^{\dagger}$CLAPSpeech-style).
  Neither retrieval measure ranks the systems.}
  \label{fig:scatter}
\end{figure}

No. The two variants with the highest prosodic sensitivity, 2+3 and 1+2+3,
differ by about 0.01--0.02 on all retrieval measures, yet they give the
second-best and the worst synthesis respectively (WER 23.4\% vs.\ 42.9\%,
IntMOS 2.13 vs.\ 1.55). R@1-diff does not rank systems either: Stage~2 alone has
the highest value but a WER of 29.0\%. In-batch retrieval measures whether
matched text and audio embeddings are nearest neighbours; it says nothing about
whether the embedding sequence is a smooth, consistent conditioning signal for a
generative decoder. Fig.~\ref{fig:scatter} shows the lack of a usable
relationship over all six variants: the Spearman rank correlation with WER is
$-0.14$ for R@1-sim and $-0.37$ for R@1-diff, and with IntMOS $0.14$ and $0.37$;
with six systems none of these is distinguishable from zero. Retrieval is still
useful as a \emph{necessary} check, since the two variants without Stage~2 are
poor on both R@1-diff and synthesis, but it cannot rank encoders that pass it.
Encoder recipes therefore have to be selected by training the TTS model.

\subsection{RQ3: does same-phoneme refinement help?}
\label{ssec:rq3}
Stage~3 behaves as intended in embedding space: appended to 1+2 it raises
R@1-sim (0.746 to 0.770). It also lowers R@1-diff (0.926 to 0.893), and the same
loss of discrimination occurs when it is appended to Stage~2 alone (0.933 to
0.871). This is expected: in same-phoneme batches gradients flow through a
single position whose phoneme identity carries no information for the loss, so
nothing preserves the discriminative features learned in Stage~2, which are
gradually overwritten (catastrophic forgetting). Same-phoneme negatives also differ only in subtle detail and give a weaker
training signal.

The downstream effect is, however, not uniform. After 1+2, Stage~3 is
destructive (WER 17.6\% to 42.9\%, IntMOS 2.19 to 1.55, the worst system
overall); after Stage~2 alone it is mildly beneficial (WER 29.0\% to 23.4\%,
IntMOS 1.89 to 2.13, within confidence intervals). The drop in R@1-diff alone
therefore does not explain the WER, which again illustrates RQ2. What the data
support is that same-phoneme refinement never outperforms the 1+2 curriculum,
is unreliable as a final stage, and used alone, as in CLAPSpeech, gives
the second-worst system. Mixing both batch types in one stage is a natural
remedy, left for future work.

\subsection{Practical implications}
\label{ssec:practice}
The results suggest three rules for pretraining TTS text encoders. First, a contrastive stage must contain negatives of other
phonemes; same-token batches alone separate phonemes poorly (R@1-diff below
0.78) and give the least intelligible speech.
Second, the text side should be conditioned on the speaker: removing the
speaker embedding cost more retrieval accuracy than any other change.
Third, a further refinement stage should be accepted only if it improves
synthesis: here the variant with the second-highest R@1-sim synthesised worst.

\subsection{Limitations}
\label{ssec:limits}
Absolute quality is low for all variants (WER $\geq$ 17.6\%, MOS $\leq$ 2.0).
We attribute this to the combination of a small 2021 acoustic model, a batch
size of 16, and noisy in-the-wild podcast speech. To keep the comparison fair,
every run was given the same fixed budget of $10^6$ steps rather than being
trained to convergence, so Grad-TTS is likely under-trained in all conditions
alike; the anchored rating scales also yield lower
values than conventional MOS. Our conclusions thus concern the \emph{relative} effect of encoder
pretraining and remain to be confirmed with a stronger acoustic model. The CLAPSpeech reference is a re-implementation
of its phoneme-level batch construction inside our architecture, not the
original model with its word-level branch, and we did not train Grad-TTS with
its original, non-pretrained text encoder. All results are single runs on one language.
\section{Conclusion}
\label{sec:conclusion}
We asked how a TTS text encoder should be pretrained to carry prosodic
information without losing phoneme identity. In a controlled Grad-TTS
ablation, MLM followed by mixed-phoneme contrastive
alignment with speaker conditioning was the best recipe, nearly halving the WER
of CLAPSpeech-style same-phoneme training and significantly improving
intonation ratings. Same-phoneme refinement improved same-phoneme retrieval but
eroded phoneme discrimination and, appended to the best curriculum, produced the
worst synthesis. Retrieval accuracy did not predict synthesis quality, so
pretraining recipes must be validated by generation. Future work will mix
both batch types in one stage and use a stronger acoustic model.


\section{Acknowledgments}
The authors used Claude Fable~5.1 (Anthropic) for language polishing and to
prepare the code of Figs.~1--3.

\section{Compliance with Ethical Standards}
The listening tests followed the principles of the Declaration of Helsinki.
The 48 participants were adult volunteers who gave informed consent; only
anonymous ratings were collected. The authors' organisations do not require
formal ethics committee review for this type of study.
The authors declare no conflicts of interest.

\bibliographystyle{IEEEbib}
\bibliography{ref_icassp}

\end{document}